# Brillouin Light Scattering: Applications in Biomedical Sciences

Francesca Palombo*[†] and Daniele Fioretto[‡]

[†] School of Physics and Astronomy, University of Exeter, Stocker Road, EX4 4QL Exeter, UK

[‡] Department of Physics and Geology, University of Perugia, via Alessandro Pascoli, I-06123 Perugia, Italy

**ABSTRACT:** Brillouin spectroscopy and imaging are emerging techniques in analytical science, biophotonics and biomedicine. They are based on Brillouin light scattering from acoustic waves or *phonons* in the GHz range, providing a nondestructive contactless probe of the mechanics on a micro-scale. Novel approaches and applications of these techniques to the field of biomedical sciences are discussed highlighting the theoretical foundations and experimental methods that have been developed to date, acknowledging that this is a fast moving field and a comprehensive account of the relevant literature is critically assessed here.

## 1. INTRODUCTION

Latest advances in label-free chemically specific imaging methods based on light–matter interaction are making a transformation in our approach to study health and disease in cells and tissues, enabling the identification of their molecular make-up with high spatial resolution. Biological and clinical samples are complex mixtures of molecules, ions and radicals and more or less organised structures, with hierarchies determining the function of each vital constituent. These building blocks are also dynamic, as they need to adapt their structure to enable a particular function – this is especially the case of proteins e.g. enzymes in the body. We have at our disposal a host of tools based on analytical devices with demonstrated applications for *in vivo* testing in biological and clinical settings. Methods based on vibrational spectroscopy such as IR absorption and Raman scattering are widely applied in biomedical studies aimed at obtaining chemical images (at diffraction-limited resolution) or individual spectra that are truly chemical fingerprints of a sample material.

Within the class of analytical techniques, Brillouin spectroscopy, which probes the mechanical and viscous properties of matter, has traditionally been applied in soft matter and material science studies, mainly in homogeneous and at-equilibrium systems. Only recently, advances in confocal micro-Brillouin light scattering (BLS) instrumentation and fast spectral analysis based on the use of single etalons have noticeably improved both spatial resolution and acquisition times. This is making it feasible to apply Brillouin spectroscopy to heterogeneous materials such as biomedical samples and out-of-equilibrium systems, thus opening the way to live cell imaging, *in vivo* tissue imaging and rapid monitoring of tissue biomechanics.

When an optical beam impinges on matter, there is a range of effects that occur, namely multiple scattering of light that can be exploited to investigate the material properties. Three types of interaction are relevant here – elastic Rayleigh scattering, inelastic Brillouin and Raman scattering. Léon Brillouin reported for the first time in 1922 the theoretical prediction of the effect by which a coherent light beam, scattered off thermally induced acoustic waves, undergoes a frequency shift equal to the frequency of the acoustic wave.[1] In 1926 Mandelstam reported a similar prediction,[2] and it was not until 1930 that the first experiment was performed by Gross.[3] BLS is the scattering of light from acoustic modes (phonons), magnetic modes (magnons). In this review, we are concerned with acoustic phonons (or optical modes in the case of Raman scattering), i.e. acoustic lattice vibrations that propagate as material density fluctuations giving rise to spontaneous BLS. Time-dependent density fluctuations result in periodic changes in the material's refractive index, which in turn acts as a diffraction grating for the incoming light. As the sound wave propagates through the medium, light will undergo a change in frequency due to Doppler shift.

Brillouin spectroscopy measures these fluctuations in dielectric constant that probe elastic properties on a micro-scale. In the ideal case of purely elastic media, i.e. with negligible attenuation of elastic waves, a typical Brillouin spectrum presents an intense central peak due to Rayleigh scattering and a set of equally shifted peaks, which are the Stokes and anti-Stokes parts of the Brillouin doublet with frequency shift

$$\omega_B = \pm Vq \qquad (1)$$

relative to the elastic peak, where $V$ is the acoustic wave velocity in the material and $q$ the momentum exchanged in the scattering process

$$q = 2nk_i \sin\frac{\theta}{2} \qquad (2)$$

with $n$ the refractive index of the material, $\theta$ the scattering angle and $k_i$ the wave vector of the incident light

$$k_i = \frac{2\pi}{\lambda_i}, \qquad (3)$$

where $\lambda_i$ is the incident light wavelength in vacuum. For a backscattering geometry ($\theta = 180°$) typical of micro-Brillouin applications, Eq.2 is simply

$$q = 2nk_i. \qquad (4)$$

Hence, the frequency shift of Brillouin peaks obtained from a combination of Eq. 1–4 is

$$\omega_B = \pm \frac{4\pi n}{\lambda_i} V. \qquad (5)$$

From the spectral shift, relevant information on the elastic properties of the material can be obtained, specifically the longitudinal elastic modulus at GHz frequencies

$$M = \rho V^2 = \rho \frac{\omega_B^2}{q^2}. \qquad (6)$$

This simple treatment must be generalised in the case of biomedical samples, where elastic properties are structured in a complex pattern of temporal and spatial scales, which are fundamental to determine the physiological conditions of biological matter. The various time scales of vibration and diffusion of molecules and macromolecules, i.e. the *temporal heterogeneity* of the material, require a viscoelastic treatment of Brillouin spectra. Moreover, the various spatial scales of organisation of biological matter, i.e. its *spatial heterogeneity* from single cells up to tissues and organs, require mapping at high spatial resolution by means of a micro-Brillouin approach. The aspects of temporal and spatial heterogeneity will be treated hereafter.

### 1.1. Brillouin Scattering from Viscoelastic Materials

Viscoelastic materials are characterised by frequency-dependent elastic moduli. The complex longitudinal modulus assessed by Brillouin scattering, $M^*(\omega) = M'(\omega) + iM''(\omega)$, shows a dispersion in the storage modulus $M'$ associated with a maximum of the loss modulus $M''$, at those frequencies where longitudinal acoustic phonons are coupled (relaxation process) with internal (e.g. molecular) degrees of freedom (see Figure 1). In viscoelastic materials, the analysis of Brillouin spectra can easily give access to $M^*$ at the single frequency of Brillouin peaks. In particular, Brillouin peaks can be reproduced by a damped harmonic oscillator (DHO) function[4]

$$I(\omega) = \frac{I_0}{\pi} \frac{\Gamma_B \omega_B^2}{(\omega^2 - \omega_B^2)^2 + (\Gamma_B \omega)^2} \quad (7)$$

convoluted with the instrumental function. The frequency shift $\omega_B$ and line width $\Gamma_B$ derived from fit analysis of Brillouin peaks yield the storage and loss moduli

$$M'(\omega_B) = \frac{\rho}{q^2} \omega_B^2 \quad (8a)$$

$$M''(\omega_B) = \frac{\rho}{q^2} \omega_B \Gamma_B \quad (8b)$$

where $\rho$ is the mass density of the material.

The single peak analysis is the standard data analysis method in BioBrillouin applications, the only difference between different approaches being in the function adopted for fitting the Brillouin peak. It is worth noting that, although the DHO function (Eq.7) derived from the hydrodynamic theory of density fluctuations gives a value of $\omega_B$ that does not exactly correspond to the maximum of the Brillouin peak, it is the correct value to estimate $M'$ via Eq.8a. Conversely, using a Lorentzian or other symmetric function to assess the position of the maximum of the Brillouin peak requires the fitting parameter to be corrected by a factor that increases with increasing line width $\Gamma_B$ to determine $M'$ through Eq.8a.[5]

A full-spectrum analysis of Brillouin light spectra, together with Brillouin spectra obtained in UV and X-ray regime and ultrasonic data can, in principle, considerably enlarge the frequency range for a complete viscoelastic characterisation of matter.[6] Having access to a large frequency range is especially useful in soft and biological matter, where a number of relaxation processes can affect the frequency dependence of $M^*$. On the other hand, fast scanning is required for mapping purposes, giving a practical limit to the quality of the spectra and allowing only single peak analysis and single frequency detection of the modulus through Eqs. 7 and 8. Since it is impossible to gauge the full relaxation scenario from a single spectrum, it is useful to compare frequency shifts and line widths obtained from adjacent points in micro-Brillouin maps, to infer information on structural arrest, moisture content, etc.

In this regard, there are three typical cases that are frequently encountered in BLS from biomedical samples (Figure 1).

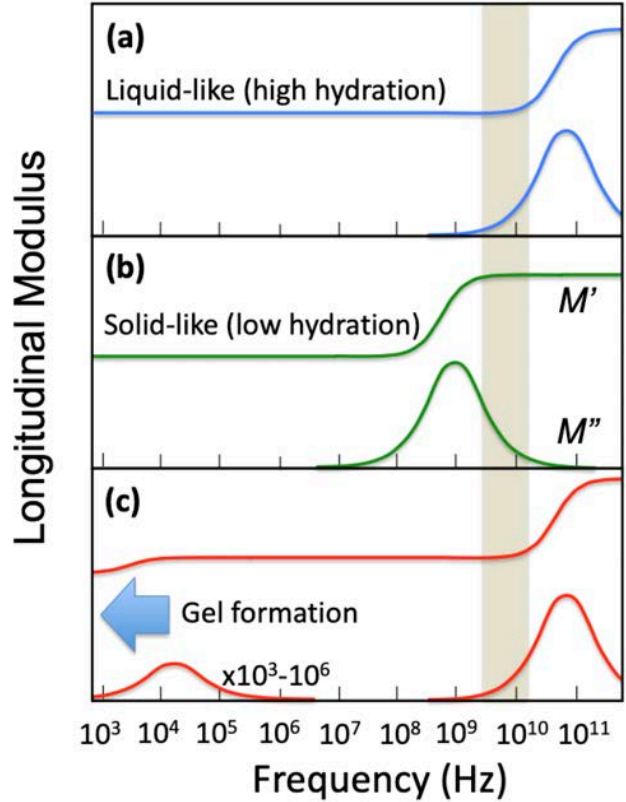

**Figure 1.** Typical dispersion curves relative to the longitudinal elastic modulus of samples with a structural relaxation rate (a) higher (liquid-like) and (b) lower (solid-like) than the Brillouin peak, usually falling within the shadowed region of the plot. The case (c) refers to a heterogeneous sample, where a small fraction of arrested state is forming within a large fraction of liquid sample, giving rise to a gel phase.

The relaxation scenario shown in Fig.1a represents highly hydrated samples, where the structural relaxation responsible for the transition from liquid- to solid-like behaviour is located at higher frequencies than the Brillouin peak. In this case, a shift of the relaxation process due, e.g., to a decrease of temperature or hydration would induce an increase in $M'(\omega_B)$ (Brillouin frequency shift, Eq.7) correlated with an increase in $M''(\omega_B)$ (line width, Eq.8). This is the case, for example, of Brillouin scattering from cytoplasm within single cells, recently reported in ref. [7]. The limiting case of essentially pure water, the so-called simple hydrodynamic condition, corresponds to $\omega\tau \ll 1$ (relaxed condition) where $M' = \rho V_0^2$, with $V_0$ the adiabatic sound velocity, and $M'' = \omega_B b$, with $b = \frac{\rho}{q^2}\Gamma_B$ the longitudinal kinematic viscosity.

By reducing hydration, the structural relaxation shifts towards lower frequencies, eventually reaching fractions of Hz or lower in the arrested state of matter. Fig.1b represents this condition, where the Brillouin peak is located at higher frequencies than the structural relaxation. In this case, a further shift of the relaxation induced e.g. by a local reduction of hydration level would induce an increase in $M'(\omega_B)$ correlated with a decrease in $M''(\omega_B)$, opposite to the case of highly hydrated samples. This is the case, e.g., of Brillouin scattering from *ex vivo* dry cartilage,[8] amyloid plaques[9] or Barrett's oesophagus.[10] The intermediate condition, when relaxation rate is comparable with Brillouin frequency shift, is the most complex, though it contains a great deal of potential. In fact, by changing a control parameter, namely the temperature, one can observe a maximum in line width. In that condition, the Brillouin frequency shift gives direct access to an important parameter of molecular dynamics, i.e. the relaxation time $\tau = 1/\omega_B$.[11]

The third case depicted in Fig.1c is also quite common in biological matter. In fact, it refers to the coexistence of both solid- and liquid-like regions in close contact to one another within a sample, in continuous sol-gel transformation.[12,13] This is a key feature for the coexistence of physiological liquid conditions (necessary for molecular trafficking, DNA replication, etc.) and of adequate mechanical resistance in cells, tissues and organs. The simplest models for this state of matter are hydrogels, where a low fraction of cross-linked macromolecules can sustain the liquid phase. The very simplified sketch in Fig.1c shows the effect of the polymeric part as the low frequency relaxation, with a diverging relaxation time that is responsible for the diverging static viscosity and for the development of recovery for shear stress, i.e. the development of a solid-like floppy shape for the sample. Conversely, the liquid fraction of the sample is responsible for the high frequency relaxation, close to the Brillouin frequency shift. The biphasic viscoelastic behaviour is quite general in liquid matter close to the gelation point, and is even responsible for the lambda transition in liquid sulfur.[14] In biological samples, quasi-static measurements are more sensitive to the rheological properties of the polymeric scaffold, whilst Brillouin scattering is more sensitive to local viscosity, i.e. to the dynamics of the liquid-like phase. In fact, the shear modulus $G$ or Young's modulus $E = 9KG/(3K + G)$, where $K$ is the bulk modulus, are usually revealed in low frequency experiments. The value of $G$ and, as a consequence, of $E$ is zero in the liquid phase, and increases up to kPa–MPa at the sol-gel transition. Conversely, the longitudinal modulus detected by Brillouin scattering $M = K + (4/3)G$ includes the contribution of the adiabatic bulk modulus that is of the order of GPa even in water. As a consequence, $M$ is less sensitive than $G$ or $E$ to the sol-gel transition and to the peculiar power law behaviour of $G'$ and $G''$ reported for biological matter, usually attributed to the cytoskeleton dynamics in living cells[15] and to its soft glassy behaviour.[16-18] In the GHz regime of the longitudinal modulus accessed by Brillouin scattering, the most promising (and less explored) potential of Brillouin micro-spectroscopy is in revealing both static ($M'$) and dynamic ($M''$) properties of the liquid fraction of biological samples, the high frequency portion of the relaxation pattern depicted in Fig. 1c. In fact, the factor $\frac{\rho}{q^2}\Gamma_B$ in Eq.8b is the so-called apparent longitudinal kinematic viscosity. Here "apparent" means that it is not related to the static viscosity but to its value at the frequency of the Brillouin peak.[19,20] We can expect this viscosity and its spatial fluctuations to correlate with viscosity and diffusion coefficients measured by local sub-micrometric rheological probes such as those employed by fluorescence correlation spectroscopy (FCS)[21] and FCS imaging.[22] An experimental effort in this direction would be motivated by the foreseen possibility of employing Brillouin micro-spectroscopy as a sub-micrometric probe of those friction coefficients that are relevant for the molecular trafficking inside cells and tissues.

It is worth pointing out that here we have painted a very simplified scenario, in order to comment on some very general behaviour. More realistic pictures should include multiple-relaxation patterns, which are typical of complex matter structured at different temporal and spatial levels.

1.2. Brillouin Scattering from Heterogeneous Materials

The spatial resolution of micro-BLS is dominated by the wavelength and attenuation of acoustic modes. In fact, an acoustic mode gives the average viscoelastic properties over its path length that is typically of the order of a micrometre in biological matter. For a reliable interpretation of the results, the multiscale heterogeneity of biological matter must be properly taken into account, since it profoundly affects both light and acoustic propagation into the sample, giving rise to relevant spectral features such as a mixing of homogeneous and heterogeneous line broadening, and multiple-scattering effects.

For this, we need to be aware of the presence of various spatial scales; three of them are relevant to micro-Brillouin measurements (Figure 2).[9] The shortest spatial scale of ca. 200 nm ($L_1$) is the typical wavelength of acoustic modes. An intermediate spatial scale of the order of 1 μm ($L_2$) is related to the attenuation of acoustic modes. The largest spatial scale of ca. 10 μm ($L_3$) is the size of the scattering volume and depends on the optical setup. An in-depth investigation of the various spatial scales involved in micro-Brillouin measurements has been given by Mattana et al.[9] Here we recall that the easiest condition to be analysed is that of a homogeneous sample, or a sample with heterogeneities much smaller than $L_1$ within the scattering volume. This is also the case of biopolymers such as collagen, the most abundant protein in mammalian tissues, which is made of triple helix molecules, each of them with a diameter of 1.6 nm and a length of 300 nm. The length scale is comparable to the wavelength of an acoustic phonon, but the diameter is so small that Brillouin scattering applied to collagen will probe an average elasticity (Voigt-Reuss model[23,24]). Conversely, micrometric large structures with mechanical heterogeneity scales in the range between $L_1$ and $L_2$ are the most complex to be analysed. This is, e.g., the case of spider silk, where the single fibre can act as mechanical waveguide, giving rise to non-trivial patterns of eigenmodes.[25] Finally, we can have heterogeneities larger than $L_2$ and smaller than $L_3$. Here $L_3$ is essentially defined by the spatial resolution of the microscopy system. At the diffraction limit, the acoustic wave vector is also at limit, so in viscoelastic materials there is no shorter scale than that probed by the incoming radiation. This is in terms of lateral resolution, whilst axial resolution is such that phonons are usually amply accommodated within this scale.

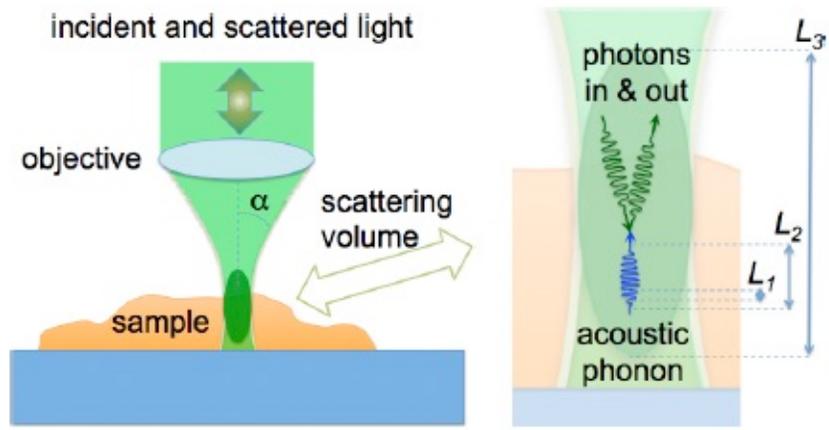

**Figure 2.** Schematic diagram of a micro-Brillouin measurement. $L_1$, $L_2$ and $L_3$ denote the relevant spatial scales in the light interaction with an acoustic phonon within the scattering volume. Figure reproduced from: ref. [9], World Scientific.

In case of heterogeneities larger than $L_2$ and smaller than $L_3$, we will have multiple peaks in the Brillouin spectrum, one for each of the mechanically homogeneous regions within the scattering volume. When the distance in frequency of these spectral features is large enough with respect to the resolving power of the spectrometer, such as in the case of cytoplasm vs. buffer in live cells[7] or of epoxy matrix vs. embedded fibre,[26] the different contributions can be extracted from the spectrum by means of an appropriate fitting routine. More complex is the case of spectral contributions very close the one another, which gives rise to a heterogeneous broadening of the Brillouin line. However, we have recently shown that also in these cases the principal components of the spectrum can be attempted by means of an hyperspectral analysis of 2D Brillouin maps.[27]

Bearing in mind these characteristic time and length scales in relation to the biomedical material, we can see that an aspect of interest concerns the temporal and spatial scales of the Brillouin measurements, which depend on the instrumentation employed.

## 2. INSTRUMENTATION – STATE OF THE ART

Brillouin microspectroscopy is capable of high spatial (~1 μm) and spectral (~$10^6$ Hz) resolution, and relies upon the non-trivial combination of two instruments: a high-resolution spectrometer and a probe, which can be a confocal microscope or a fibre-optic device or else. An overview of the most recent ad-vances in these technologies, that is at the basis of the increasing success of micro-BLS and of its use for different applications, is reported here.

Confocal laser scanning microscopy is a well-established technology based on the use of large numerical aperture (NA) objectives and narrow spatial filters. 3-D mapping of biological samples by confocal microscopy is typically based on fluorescent probes. These emit intense light at frequencies that are far lower than the excitation source, so that a very effective spectral filtering can easily be achieved, producing large field of view (FOV) images very rapidly. On the other hand, observing Brillouin signals by confocal microscopy is hampered by the low scattering cross-section of spontaneous BLS from bulk phonons[28] and by the very small frequency shift of the scattered light ($10^{10}$ Hz, over $10^{15}$ Hz of the excitation source), demanding for spectral filters of very high contrast and resolution. Tandem multipass Fabry-Pérot interferometers (TFPI) are especially suited for this purpose, as they have unparalleled contrast and spectral resolution; however, even their most recent version suffers from limited acquisition rates (~1 spectrum/s) in poorly scattering media. Virtually Imaged Phased Array (VIPA) spectrometers have contributed to the development of BLS microscopy by achieving spectral dispersion via single large angular dispersive etalons with mirror surfaces,[29] reducing by a factor of 100-1000 the acquisition time of a single spectrum. Unfortunately, this type of spectrometers is hampered by a low contrast (~55 dB for a traditional double-VIPA[30]) compared to the TFPI (>150 dB[26]) so that it has typically been used for low-turbidity samples, e.g. cells in aqueous medium or highly hydrated materials.

2.1. Fabry-Pérot Interferometer

Advances in tandem Fabry-Pérot interferometers, namely TFP-1 and TFP-2 HC spectrometers,[26] have enabled applications where a strong elastic light component is mixed with the Brillouin signal, or when it is necessary to perform measurements at different free spectral ranges (FSR). In these spectrometers, two sets of parallel mirrors are scanned by means of piezoelectric transducers: the spacing between mirrors determines the FSR, which is tunable, whilst high contrast is obtained by passing the transmitted peaks three times across each of the two FPs in a tandem arrangement (Figure 3).

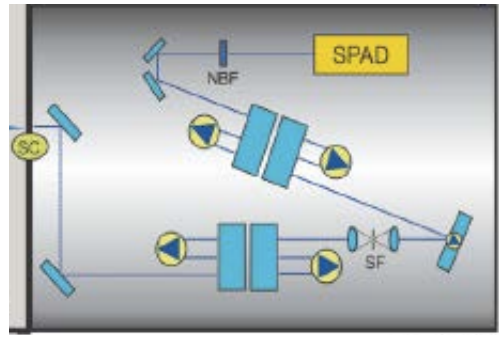

**Figure 3.** Schematic diagram of a high-contrast Tandem Fabry-Pérot interferometer (TFP-2 HC). It includes a spatial filter (SF), a narrow band pass filter (NBF), and a Single-Photon Avalanche Photodiode (SPAD). Figure reproduced from: ref. [26], American Physical Society.

The contrast is unprecedented, >$10^{15}$ (or 150 dB) in the latest development, and the scanning mode is the only price to pay. This is not an issue in systems at equilibrium, but it can be unfavourable in living systems that move and hence require fast acquisition rates. VIPA spectrometers may be the preferred option in those cases, when high contrast is not a stringent requirement. It is worth noting that often a rapid mapping realised at the expense of laser radiation power (high power density delivered to the sample) can be detrimental to live cells, tissues and organisms, hence it is preferable to extend the acquisition time while working at low laser power or low NA.

Brillouin spectroscopy performed in backscattering geometry with a high NA objective or in fibre optics need accounting for the fact that the exchanged momentum $q$ is angle-dependent (Eq.2). Hence, light accepted at higher angles results in asymmetric broadening of the Brillouin signal, as explained by Mattana et al.[7] Measurements performed with other scattering geometries are even more impacted by this effect. The most compelling examples of line shape analysis to date are those performed on FP-Brillouin spectra,[7,27,31] since the increased spectral resolution make it possible to assess the "true" line width, devoid of the contribution from the instrumental function, and hence to accurately determine the acoustic modes' attenuation parameters. This has recently been demonstrated in the first hyperspectral Brillouin data analysis based on multivariate statistics[27] (see below).

2.2. Virtually Imaged Phased Array

Since the introduction of Brillouin imaging,[32] Scarcelli and Yun have made the most compelling instrument development in VIPA-Brillouin microscopy applied to the life sciences.[30,33-35] Multistage VIPA spectrometers are the equivalent to the TFPI, and achieve higher contrast than single etalons.[36] Parallel efforts to enhance the contrast of VIPA spectrometers have been made using a range of techniques including gas-chamber filtering,[37] apodization,[35] destructive interference,[38] narrow band-pass filtering,[39] and spectral coronagraphy.[40]

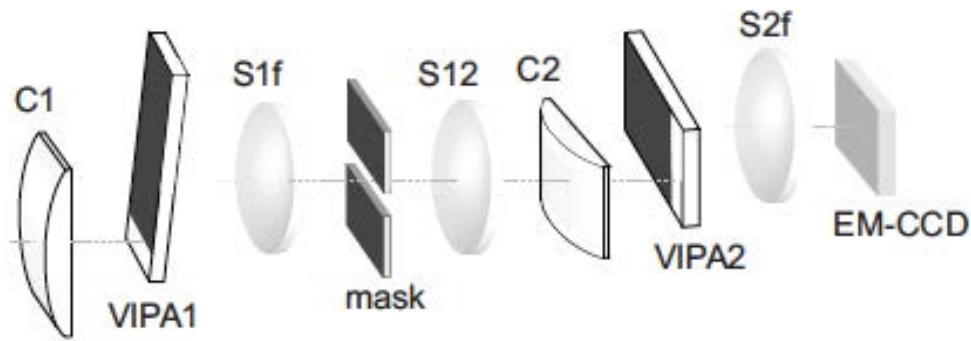

**Figure 4.** Schematic diagram of a two-stage VIPA spectrometer. It includes two etalons (VIPA1/2), two cylindrical lenses (C1/2; 200 mm focal length), three spherical lenses (S1f, S12 and S2f; 200 mm focal length), and an Electron Multiplying Charge-Coupled Device (EM-CCD) camera. Figure adapted from: ref. [36], Optical Society of America.

Attempts to improve acquisition rates include line scanning with a single VIPA.[41] This enables hundreds of points in a sample to be measured simultaneously using line-scanning parallel detection, which reduces the acquisition time of hyperspectral maps from hours to tens of seconds at the expense of the contrast (~30 dB). Rapid high-sensitivity approaches are those based on coherent or stimulated Brillouin scattering,[42] (see Section 5.2) which are very promising, despite the complexity and costs of instrumentation.

Acquisition rate is usually a trade-off with the laser output power, which for biomedical applications needs to be carefully tailored so that any detrimental effects to the sample (e.g. cytotoxicity due to exposure to high energy laser radiation) are reduced. Lowering the NA of the microscope objective can be used to reduce the laser power density. Whilst this reduces the achievable spatial resolution, it also reduces the spectral broadening,[43] which is favourable for an accurate line shape analysis.

# 3. ADVANCES IN BRILLOUIN SCATTERING METHODS

3.1. Spectroscopy and Imaging

The first studies using Brillouin spectroscopy on biologically relevant samples date back to the 1970s. The first study on biopolymers of which we are aware of by Bedborough & Jackson[44] has applied Brillouin spectroscopy to collagen gels at various concentrations. The most prominent effect of changing water content in collagen gels is not in the velocity of sound (i.e. frequency of the Brillouin lines) but rather in the line widths. This advocates the application of a full line shape analysis, beyond what shown in the limit of very high hydration in non-biopolymer gels,[45] for a complete description of viscoelasticity by Brillouin scattering. The subsequent work by Zhao & Vanderwal applies a viscoelastic model to describe Brillouin data from collagen gels at various concentrations and temperatures.[46] It concludes that "the storage moduli of fluid and network both contribute to the longitudinal storage modulus of the system…" and "the network modulus increases with concentration".

Harley et al.[47] have measured a microscopic elastic modulus in both type-I collagen fibres and muscle proteins which they have related to the forces of hydrogen bonding along the protein backbone. Later studies by Randall and Vaughan have reported the acoustic wave velocity and attenuation in rat tail collagen[48] and rabbit psoas muscle fibres,[49] and the elastic moduli of refractive tissues of the eye.[50] Cusack & Miller[51] determined the angle dependence of both lateral and transverse modes in wet and dry collagen fibres. They fitted their data to the theoretical model of wave propagation within an elastic medium with hexagonal symmetry[52] to determine the five components of the elasticity tensor and thence the axial and transverse elastic moduli and the shear modulus of collagen. This powerful approach is based on the use of a *platelet* geometry ($\theta = 90°$) with reflective substrate, as well as on the measurement of polarization-resolved Brillouin spectra (which requires a high contrast spectrometer for revealing transverse modes). As in the previous study, the moduli derived were much larger than those determined at lower frequency by macroscopic mechanical testing. The authors speculated that this is possibly a viscoelastic effect attributable to the different timescales of the two measurement techniques.[51] For polymers, differences in Young's moduli of an order of magnitude between microscopic measurements and macroscopic stretching have been reported,[53] and the viscoelastic behaviour has been characterised.[54] The effect of hydration dynamics on the Brillouin spectrum of oriented DNA films, showing a coupling of DNA vibrations to a water relaxation as the main mechanism of phonon damping, has been reported.[55-57] Our works on type-I collagen and elastin fibres, articular cartilage and nuchal ligament, have gained insights into the mechanical anisotropy of fibrous proteins of the extracellular matrix (ECM) and on the effect of hydration on the Brillouin peaks.[8,58,59] It was found that water plays a prominent role in the dynamics and spectral features in Brillouin scattering that can mask mechanical anisotropy when signals are very close to the water modes. All these studies, focused on elastic properties of the ECM and connective tissue, are critical for the interpretation of Brillouin-derived quantities for whole tissue analysis.

Micro-BLS applications have shown that spatially resolved viscoelastic properties enable the study of heterogeneous biomedical systems. Extensive application of VIPA-Brillouin microscopy to the eye has been performed by Scarcelli and Yun.[33,34,60-62] Dependence of the Brillouin frequency upon corneal ageing, collagen crosslinking, hydration and degeneration has been demonstrated (*vide infra*). Parallel works by Reiß et al.[63,64] and Lepert et al.[65] have investigated the mechanical properties of the eye lens and cornea, respectively. Whilst the eye has obvious advantages in that it is made of essentially transparent tissue, other tissue types have also been investigated for other BLS applications.

VIPA-Brillouin–Raman microscopy applied to adipose tissue from obese rats fed a fat-enhanced diet shows greater Brillouin shift than adipose tissue from normal control rats, and the shift is greater for brown fat than for white adipose.[66]

The elastic properties of the cortical bone of bovine femur have been investigated by FP-Brillouin spectroscopy with reflection-induced scattering (or platelet) geometry.[67] Acoustic wave velocities for cortical bone after decalcification with EDTA are significantly lower than for bone before decalcification (3.3 vs. 5.1 km/s), and there exists a relationship between wave velocity and hydroxyapatite content. From the same team, comparative analysis by BLS and scanning acoustic microscopy has been conducted on trabecular bone to assess elastic anisotropy on a microscale.[68] Longitudinal sections of a bovine femur were sliced and polished, prior to acoustic impedance and platelet BLS measurements along two orthogonal directions (in-plane and out-of-plane). Results highlight the importance to map local values of mass density in order to gain accurate elastic anisotropy ratio from both techniques. A further study of anisotropic elasticity of a single rod-type trabecula derives the in-plane acoustic wave velocity as a function of the angle to the rod axis, showing slight uniaxial anisotropy (Figure 5).[69]

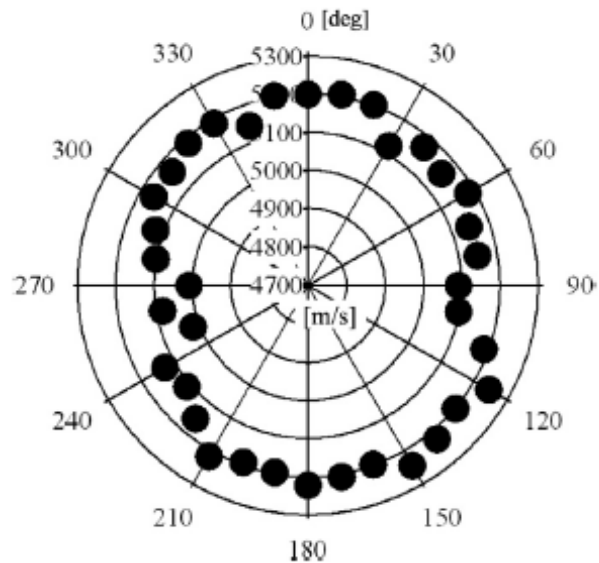

**Figure 5.** In-plane acoustic wave velocity anisotropy of a trabecular rod measured by Brillouin spectroscopy. 0° and 180° indicate the orientation of the trabecula. Figure reproduced from: ref. [69], Elsevier.

Wave velocities of rod-like trabeculae are higher than for plate-like trabeculae, indicating that elastic properties are dependent upon the trabecular type, length and orientation.[70]

3.2. Cell Imaging

Owing to its characteristics of being contactless and mechanical loading-free, Brillouin microscopy is especially amenable to applications in cell mechanics and mechanobiology. Single cell imaging studies have been reported by Scarcelli et al.,[35] Antonacci & Braakman,[71] Meng et al.[72] and Mattana et al.[7] Figure 6 shows Brillouin microscopy images of an NIH 3T3 mouse fibroblast cell before and after hyperosmotic shock (induced by adding 50 mM sucrose). The images based on the Brillouin frequency shift indicate a sizeable increase in frequency throughout the cell that can be related to an increase in longitudinal elastic modulus.[35] Note that water has a Brillouin peak at 7.5 GHz and a bulk modulus of 2.2 GPa; it accounts for 70% or more of total cell mass.

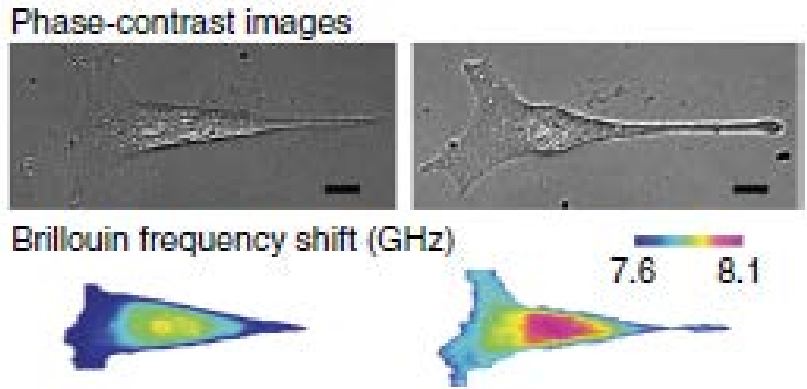

**Figure 6.** Brillouin microscopy (lower panels) and co-registered phase-contrast microscopy (upper panels) images of an NIH 3T3 mouse fibroblast cell before and after hyperosmotic shock. Scale bars, 10 μm. Figure reproduced from: ref. [35], Nature America Inc.

NIH 3T3 fibroblast cells were also investigated by means of on-chip Brillouin spectroscopy to evaluate the effect of chromatin decondensation on the nucleus response.[73] TSA-induced chromatin decondensation is known to reduce nuclear stiffness: this has been demonstrated through a reduction in the Brillouin frequency shift of the nucleus between treated and control cells.

Subcellular resolution of VIPA-Brillouin microscopy has been demonstrated in ref. [71], with results showing greater longitudinal elastic modulus of nucleoli compared to both the nuclear envelope and cytoplasm of a single endothelial cell *in vitro* (Figure 7).

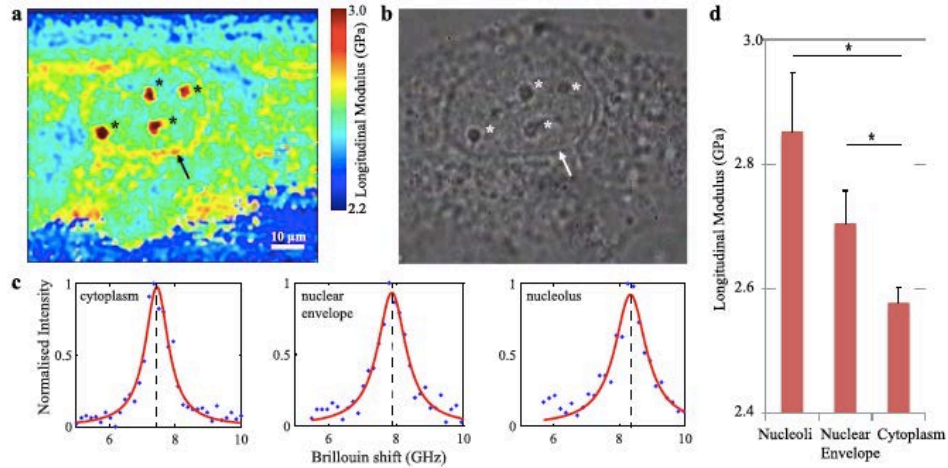

**Figure 7.** Brillouin and phase contrast microscopy (upper panels) images of a human umbilical vein endothelial cell taken at 100x magnification. Scale bar, 10 μm. Representative Brillouin microscopy (lower panels) spectra of the cytoplasm, nuclear envelope, and nucleoli. Bar plot (right panel) of the longitudinal modulus (mean±SEM) of nucleoli, nuclear envelope, and cytoplasm. Figure reproduced from: ref. [71], Nature America Inc.

Also, a marked reduction in modulus has been observed by treating a PAEC cell with latrunculin-A, hence indicating that Brillouin microscopy is a sensitive tool to investigate cellular stiffness in response to external stimuli.[71]

Karampatzakis et al. used VIPA-Brillouin microscopy to investigate the mechanical properties of *Pseudomonas aeruginosa* biofilm colonies growing in a flow cell.[74] Brillouin shifts at different depths show no correlation with colony size or flow velocity, whilst heterogeneous distributions of Brillouin shift across a single colony are apparent.

3.3. Spectral Pathology

The applications of Brillouin scattering in medical imaging have been significant for ophthalmology. Scarcelli and Yun have made compelling advances in spectral pathology related to keratoconus, with a clinical trial that has enrolled more than 200 individuals to date.

The first *in vivo* human eye application was realised with a Brillouin optical scanner, using a cw laser at 780 nm and 0.7 mW power delivered to the eye through a low-NA objective lens.[34] Brillouin measurements acquired by a two-stage VIPA spectrometer and an EM-CCD camera, with 0.4 s acquisition time, show varying Brillouin shift across depth profiles of the cornea and lens.

Maps of Brillouin shift of *in vivo* normal (n = 7) and keratoconus (n = 6) corneas show that normal corneas have relatively uniform anterior Brillouin shifts in the central region, whereas keratoconic corneas present strong spatial variations in Brillouin shifts (Figure 8).[75]

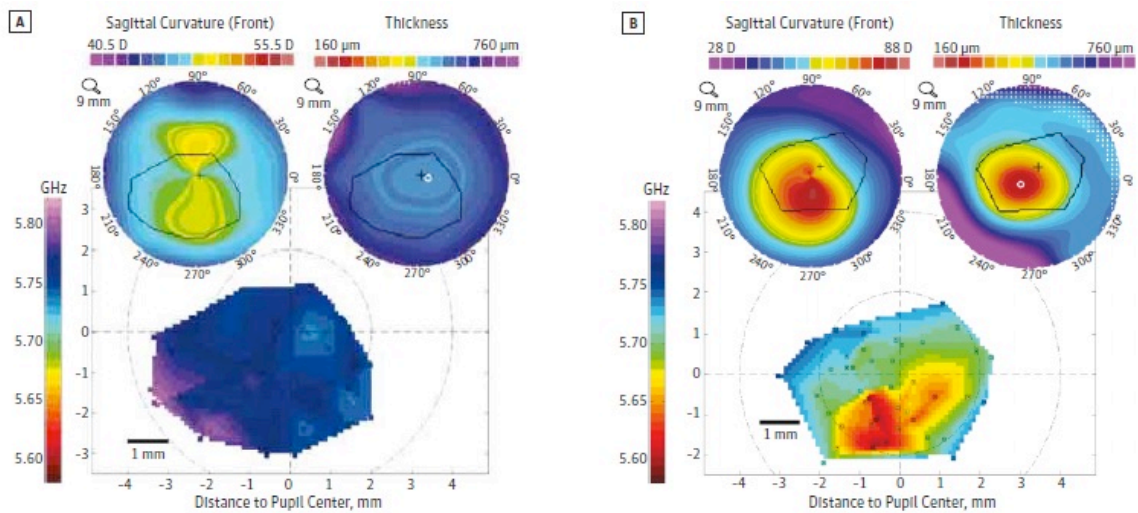

**Figure 8.** (Upper panels) Representative maps of the mean anterior Brillouin shift for a 53-year-old with normal corneas (left) and a 40-year-old patient with advanced keratoconus (right). Insets are the respective curvature and pachymetry maps with outlined Brillouin-scanned areas. Figure reproduced from: ref. [75], American Medical Association.

Our work on human tissue biopsies of epithelial tissue in Barrett's oesophagus related the viscoelastic nature of the material to the presence of hydration water in the ECM, and of ECM-epithelial cell interfaces that are revealed by BLS microscopy [10,76] (see Section 4.1).

Besides cancer-related studies, another medically relevant application has been reported for atherosclerosis in *ex vivo* histological sections of carotid artery from a mouse model.[77] Figure 9 displays two cross-sections, from instrumented and control vessels, imaged with VIPA-Brillouin microscopy.

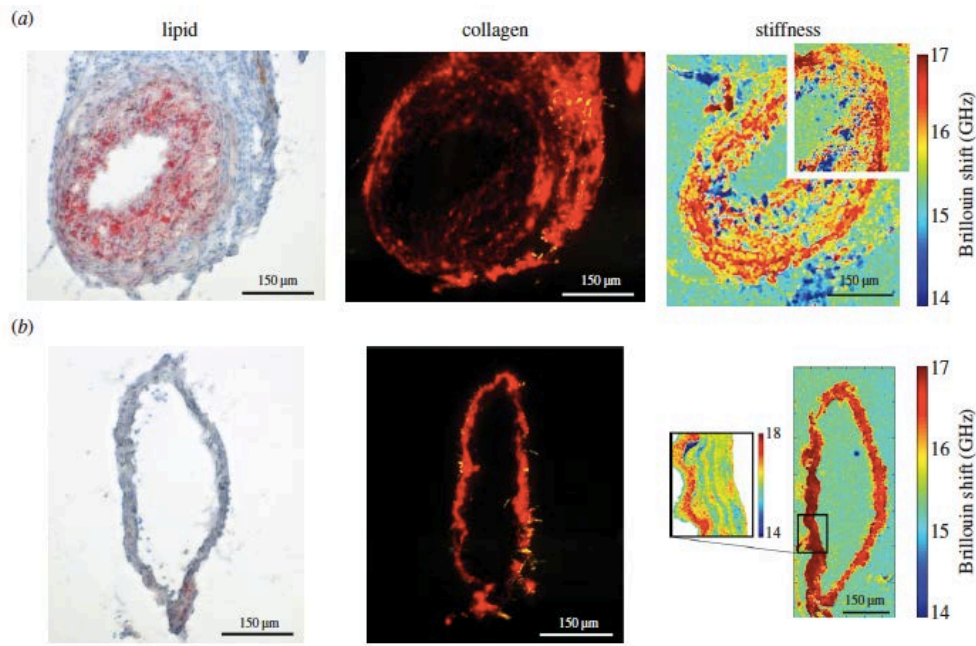

**Figure 9.** Light microscopy (left), polarized light microscopy (centre) and Brillouin microscopy (right) images of a cross-section from an instrumented (upper panels) and control (lower panels) ApoE$^{-/-}$ mouse carotid artery. Inset is a higher resolution map. Scale bars, 150 μm. Figure reproduced from: ref. [77], Royal Society Publishing.

The mean frequency shift across the control sections (17.2 GHz) is larger than within the atherosclerotic plaque (15.8 GHz), and this can be explained based on tissue composition (larger shift – collagen-rich tissue, smaller shift – lipid-rich tissue).

Background-deflection VIPA-Brillouin microscopy in combination with fluorescence imaging has been applied to intracellular stress granules containing amyotrophic lateral sclerosis (ALS) mutant FUS protein in fixed HeLa cells[78] (see Section 4.2).

Other applications of Brillouin-Raman microscopy to neurodegenerative diseases have also been demonstrated (Alzheimer's disease; see Section 4.1.3).

A promising application of Brillouin spectroscopy for pathology screening is to biofluids or liquid biopsies, as shown by Yakovlev and co-workers with a cerebrospinal fluid (CSF) model of bacterial meningitis.[79]

## 4. MULTIMODAL MICROSCOPY

Brillouin microscopy originates from the hybridization of Brillouin spectroscopy and confocal microscopy. It is thus tempting to improve the setup towards a multimodal microscopy and spectroscopic imaging. The benefit of multimodal imaging is very high when complementary spectroscopic techniques are combined and a correlation is established between elastic and structural or molecular properties of the sample.

4.1. Brillouin-Raman Microscopy

Coupling Raman with Brillouin microspectroscopy enables chemical specificity to be assigned to mechanical properties of a sample. Raman microscopy is capable of label-free chemical mapping of bio-materials. It has subcellular resolution and has made progress towards translation to clinical theranostics.[80] Brillouin and Raman scattering acquired from similar spatial scales produce an ideal combination for multimodal mapping of biomedical specimens, which has firstly been applied to *ex vivo* biopsy sections of epithelial tissue in Barrett's oesophagus (Figure 10).[10,76]

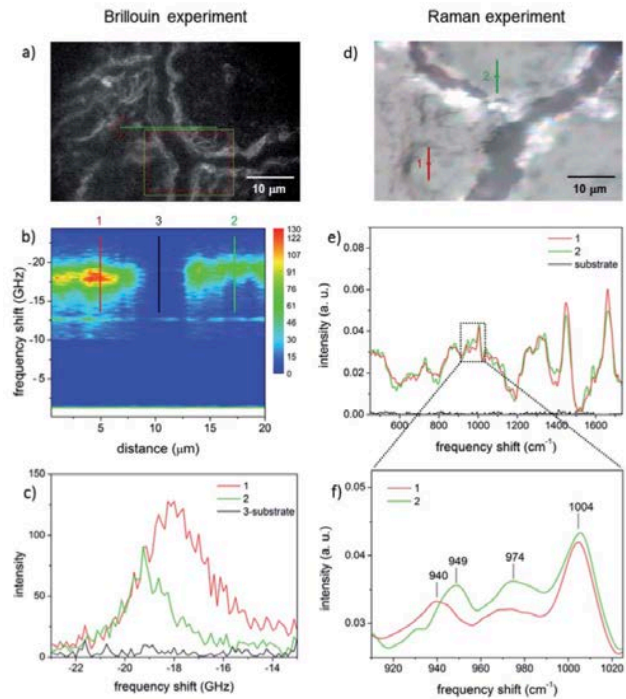

**Figure 10.** Dark-field and light microscopy (upper panels) images of a histological section of Barrett's oesophagus tissue biopsy. Scale bars, 10 μm. Brillouin intensity map and representative Raman spectra of (1-red) connective tissue and (2-green) epithelial cells (mid panels). Brillouin and Raman (lower panels) spectra extracted from the maps. Figure reproduced from: ref. [10], Royal Society of Chemistry Publishing.

Hydrated (type I) collagen fibrils forming the extracellular matrix and embedded epithelial cells were identified by Raman spectra at the gland/connective tissue interface. Contextual micro-Brillouin maps showed signatures of high scattering intensity in correspondence of low frequency shift, which was attributed to the role of bound water within collagen fibrils. In particular, the observed anti-correlation between acoustic wave velocity and attenuation suggests that hydration water has a plasticising effect on the collagen-rich connective tissue.

After understanding the potential of combined Brillouin and Raman spectroscopy, the next step was to realise a single setup for the joint measurement of both signals. This concept was "easily" realised thanks to the fact that both spectroscopies are based around inelastic light scattering, and they only differ in the probed frequency range. In fact, Raman spectroscopy concerns light scattered from vibrational molecular modes, with frequency shifts typically larger than 1 THz. The experimental layout for the combined technique can thus share the laser source and the microscope objective for focusing and detecting the scattered light. A schematic diagram of a novel FP-Brillouin–Raman microscope[26] is shown in Figure 11.

In the most efficient configuration, the scattered light is split by an ultrasteep short-pass filter, which transmits the anti-Stokes quasi-elastic scattered light to a high-resolution interferometer and reflects the Stokes deeply inelastic scattered light towards a Raman spectrometer. Setups have been realised making use of either a Fabry-Pérot interferometer, as in Fig.10,[26] or a VIPA spectrometer.[81]
1414

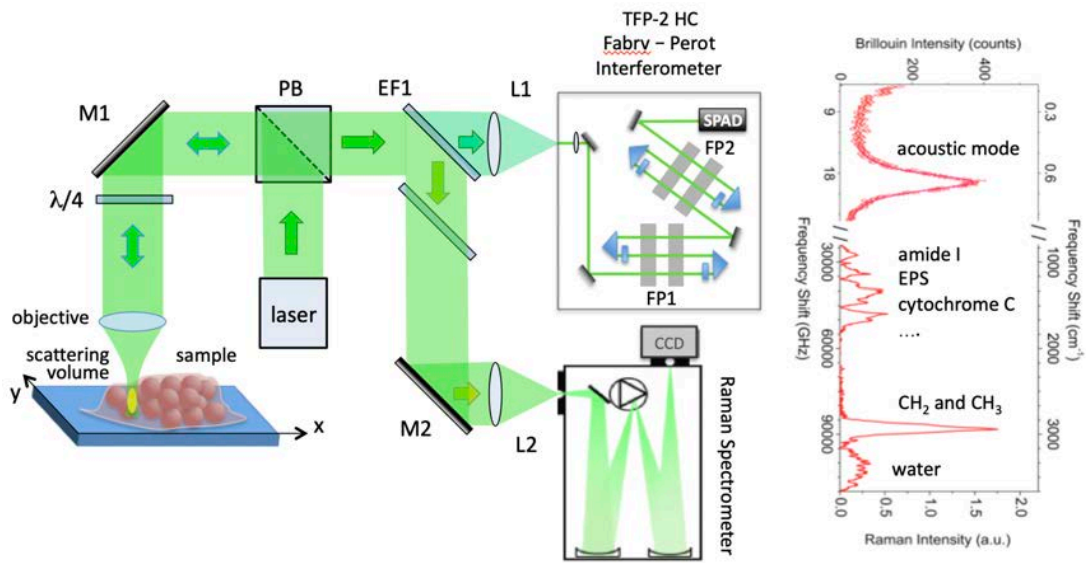

**Figure 11.** Schematic diagram of a micro-Brillouin–Raman system (upper panel). TFP-2 HC denotes a novel high-contrast tandem Fabry-Pérot interferometer. Correlative Brillouin and Raman spectra from the same location within a sample are reported (lower panel). Figure adapted from: ref. [26], American Physical Society.

In recent years, the combined Brillouin–Raman micro-spectroscopy has successfully been applied to study biological samples of increasing complexity, ranging from single cells to biofilms and animal and human tissues, as described hereafter.

4.1.1. Multimodal Mapping of Single Cells

A proof of principle that Brillouin and Raman spectroscopy can be employed for the study of the microscopic elasticity and chemical structure of individual cells has been reported for red blood cells (Figure 12).[72]

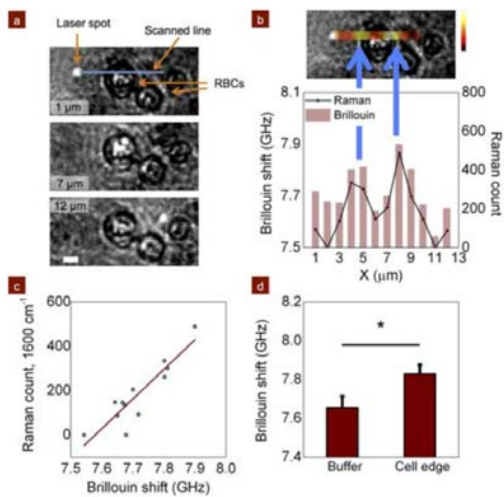

**Figure 12.** Light microscopy (upper panels) images of red blood cells. A Brillouin-Raman line scan was performed, with the bar plot showing overlaid Brillouin shift and Raman intensity data. Correlative Raman-Brillouin plot and Brillouin shift bar-plot (lower panels) for the spectra extracted from maps. Figure reproduced from: ref. [72], Wiley.

Intracellular elasticity fluctuations are identified on native RBCs, with Brillouin shifts from the cells edges being greater than those from the cell centres. The Brillouin shift positively correlates with the presence of haemoglobin revealed by Raman spectroscopy. Experimental results also suggest that the fluorophore functionalization alters the elasticity of the RBCs.

More recently, cancer live cell models[7] have been investigated at sub-micrometric resolution, correlating viscoelastic properties and chemical composition. The high content of water is a challenge, and decomposition of cell component from buffer Brillouin signal is required, as shown in Figure 13.

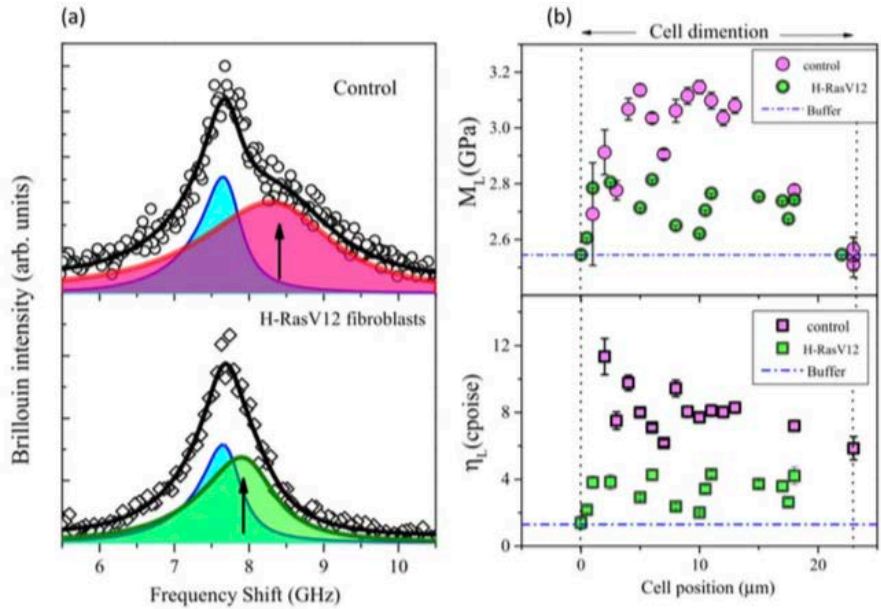

**Figure 13.** Brillouin peaks (left) from the nucleus of a control cell (NIH/3T3; upper panel) and a transfected fibroblast (H-RASV12; lower panel). Longitudinal elastic modulus and apparent viscosity (right) of the cells (pink and green symbols, respectively). Figure reproduced from: ref. [7], Nature America Inc.

After decomposition, a clear mechanical heterogeneity was demonstrated inside the cell, with an increase of 20% in the elastic modulus going from the plasma membrane to the nucleus, independently identified by Raman spectroscopy markers.

A careful Brillouin line shape analysis is even more relevant when comparing cells under physiological and pathological conditions. It has been found that, after oncogene expression, cells show an overall reduction of the elastic modulus (15%) and apparent viscosity (50%), as can be seen in Figure 13. These results can explain the invasive potential of cancer cells, since the increase in deformability enhances the squeezing ability of these cells through the extracellular matrix, favouring their dissemination and metastasis.

4.1.2. Multimodal Mapping of Biofilms

At a higher level of complexity, dry microbial biofilms have been studied to understand the microscopic reason for the survival of these primitive microbial assemblies to hostile environmental conditions.

In fact, the creation of a biofilm is one of the most important factors to guarantee the survival and dissemination of bacterial and fungal cells throughout different habitats, including those of medical and clinical environments. Dry yeast biofilms of *Candida albicans*[26] and *Candida parapsilosis*[82,83] have been studied by Brillouin–Raman microspectroscopy, exploiting the potential of mechanical mapping with chemical specificity of the multimodal approach. The morphology of a *C. albicans* biofilm grown over an aluminum substrate is apparent in Figure 14.

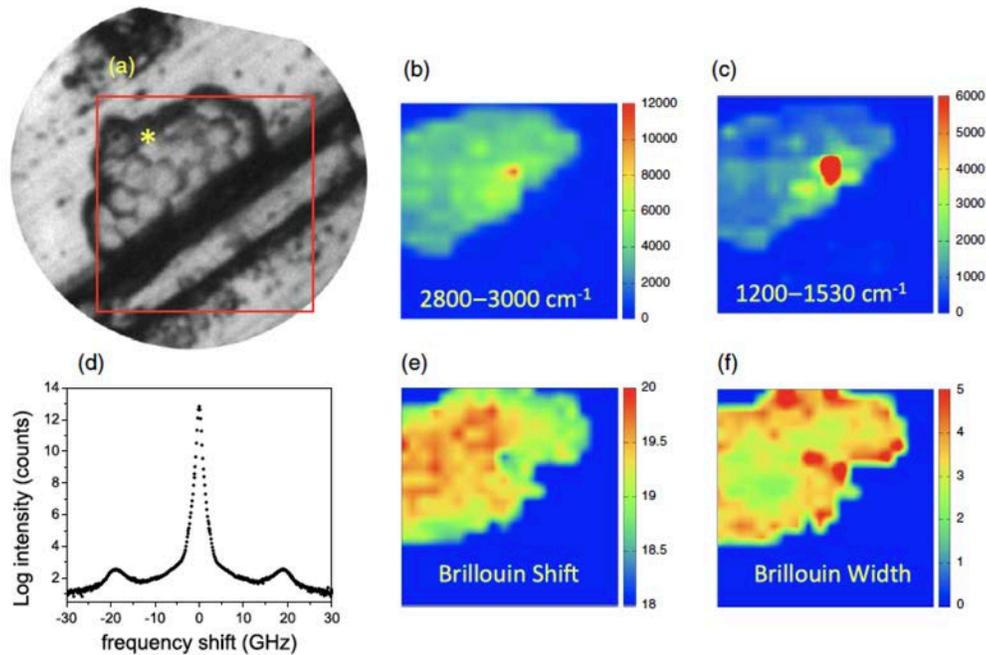

**Figure 14.** Bright-field image and Raman maps (upper panels) of a *Candida albicans* biofilm. Red box denotes the 20 μm x 20 μm area of the Brillouin–Raman map. Representative Brillouin spectrum and Brillouin maps (lower panels) of the biofilm. Figure reproduced from: ref. [26], American Physical Society.

In the central region of the map, a 6% reduction of the Brillouin frequency shift together with a 2-fold increase in the line width is visible. These features are correlated with a 3-fold increase in the intensity of Brillouin lines, suggesting that a 3-layer structure of the biofilm exists in the same region. The joint Raman investigation shows an order of magnitude increase in water content, together with an increase of Raman signals in the 1510–1750 cm$^{-1}$ region, which can be attributed to the resonant scattering from cytochrome c, a marker of cell vitality. These results (sample thickness, viscoelastic behaviour, and chemical modification) have been interpreted in terms of a buried region of the film that is favourable to the survival of *Candida* cells. In fact, it is plausible that, in the region of larger thickness, the buried cells are protected by the overlaying biofilm structure, preserving water and vital conditions. This result gives a microscopic explanation to the microbiological evidence that the biofilm acts as a structure that increases the resistance of yeasts.

4.1.3. Multimodal Mapping of Tissues

The next step in complexity is given by mechanical, structural and chemical properties of tissues. Also in this field, combined Brillouin–Raman microspectroscopy has demonstrated a great potential, which let us foresee its possible implementation as a powerful tool in histopathology laboratories.

An example is given by the recent investigation of Alzheimer's mouse brain in *ex vivo* sections of the hippocampus (Figure 15).[9]

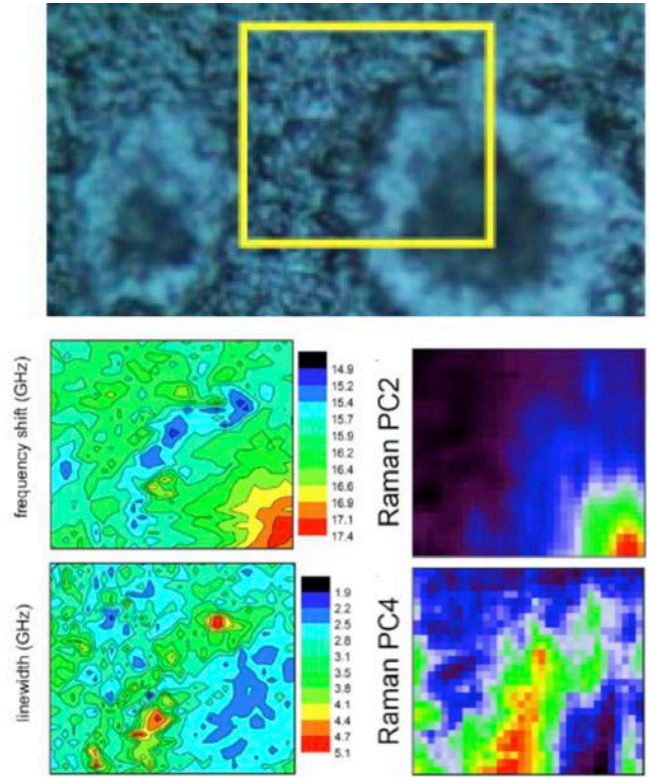

**Figure 15.** Photomicrograph and Brillouin maps (left) based on Brillouin shift (mid panel) and line width (lower panel) of an Aβ plaque portion of a transgenic mouse brain section. Yellow box denotes the 50 μm x 43 μm area of the Brillouin–Raman map. Raman maps (right) are score images obtained from Principal Component Analysis (PC2-mid panel and PC4-lower panel). Figure reproduced from: ref. [9], World Scientific.

Amyloidopathy is characterised by abnormal deposition of the amyloid beta (Aβ) peptide, giving rise to plaques with a rigid core that has successfully been contrasted by means of Brillouin microscopy (Fig.14). Dense-core plaques are surrounded by a lipid-rich layer, which is heterogeneously composed of extracellular matrix, presumably disseminated of glial cell bodies. Correlative micro-Raman analysis of the plaques shows the chemical specificity to identify the molecular origin of the biomechanical response, hence being able to relate high rigidity to the Aβ plaque core and low rigidity to the lipid halo.

In these samples, a marked heterogeneous contribution to the Brillouin line broadening has also been revealed, which is hardly decomposed via single spectra analysis. To tackle this problem, a new method has been proposed, based on an unsupervised non-negative matrix factorization procedure[27] that had already proved successful in decomposing chemical images. By this procedure, a decrease in tissue rigidity from the core through to the periphery of the plaque is revealed, with spatially distinct components that can be assigned to specific entities, as described below.

Parallel efforts to couple Raman and VIPA-Brillouin microscopy have been made by Yakovlev and co-workers.[66,72,81] In particular, they have recently applied this technique to the study of local heterogeneity in mechanical properties of nanostructured hydrogel networks.[84]

4.2. Extended Depolarized Light Scattering

Another benefit arising from coupling a Fabry-Pérot interferometer with a Raman spectrometer is that it is possible to perform extended depolarized light scattering (EDLS) experiments with a single setup.[26]

The tandem Fabry-Pérot interferometer can access a wide spectral range, from fractions to some hundreds GHz, and the Raman spectrometer can extend this range up to some tens of THz. By an analyzer that is crossed to the polarization of the laser light, we filter out the Brillouin peaks and only detect light scattered by anisotropy fluctuations, with polarization orthogonal to that of the laser. EDLS spectroscopy has been developed[85,86] to probe the vibrational and relaxation dynamics of matter that are relevant to a number of research themes in condensed matter, from hydration of biological systems[87] to anharmonicity and Boson peak in glasses and highly viscous media.[88]

The recent development of a micro-focused setup[26] gives access to subcellular structures, as shown in Figure 16, where the susceptibility spectrum of the nucleus of an NIH/3T3 murine fibroblast cell is reported.

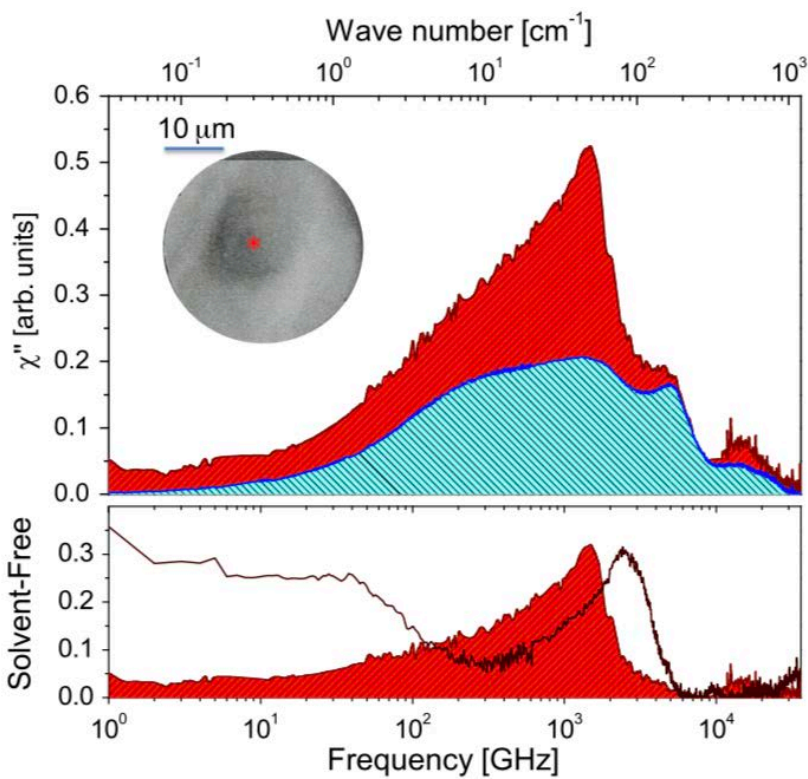

**Figure 16.** (Upper panel) Extended depolarized light scattering spectrum (represented by the susceptibility anisotropy) of the nucleus of an NIH/3T3 mouse fibroblast cell (red) and PBS solution (cyan). (Lower panel) Spectrum (red) obtained by subtracting the spectral profile of the PBS solvent from that of the cell nucleus. The solvent-free spectrum of a 100 mg/ml lysozyme aqueous solution is also reported (brown). Figure reproduced from: ref. [26], American Physical Society.

It can be seen that the spectrum of the cell nucleus is dominated by a strong vibrational contribution around 1 THz, which can be mainly attributed to the Boson peak, an ubiquitous signature of collective vibrations in disordered condensed matter.[89] It is worth noting that, after subtraction of the solvent spectrum, the residual intensity in the 10–100-GHz region can be attributed to the relaxation dynamics of hydration water, which is retarded by a factor greater than 3 relative to bulk water. It is thus easy to understand the potential of this technique in a wide range of research areas, such as understanding the hydration dynamics of biomacromolecules in a physiological state or detecting intracellular picosecond collective motions which are deemed responsible for mediating biochemical reactions and energy transport[90] and for controlling drug intercalation in DNA.[91]

## 4.3. Brillouin–Fluorescence Microscopy

Another benefit from a multimodal configuration is the possibility to improve the contrast in imaging biological matter. In fundamental biology, one is often interested in the correlation of the mechanical properties to the morphological features, which are not directly accessed by Brillouin microscopy. To overcome this problem, fluorescence emission–Brillouin scattering imaging (FBi) has been developed, a method for the parallel measurements of mechanical properties and fluorescence in living organisms.[92]

The fluorescence–Brillouin light scattering microscope was realised with excitation light from a single mode 532 nm laser, coupled into an inverted microscope frame and focused onto the sample with a high numerical aperture objective. Backscattered light from the sample is focused through a pinhole (confocal configuration), and quasi-elastically scattered light passing through a 2 nm band-pass filter is analysed by a VIPA spectrometer, whilst longer wavelength light is focused onto a photomultiplier tube for fluorescence emission measurements. The method has been tested by measuring the stiffness of plant ECMs and plant cell cytoplasm at sub-micrometre resolution in 3-D (Figure 17).

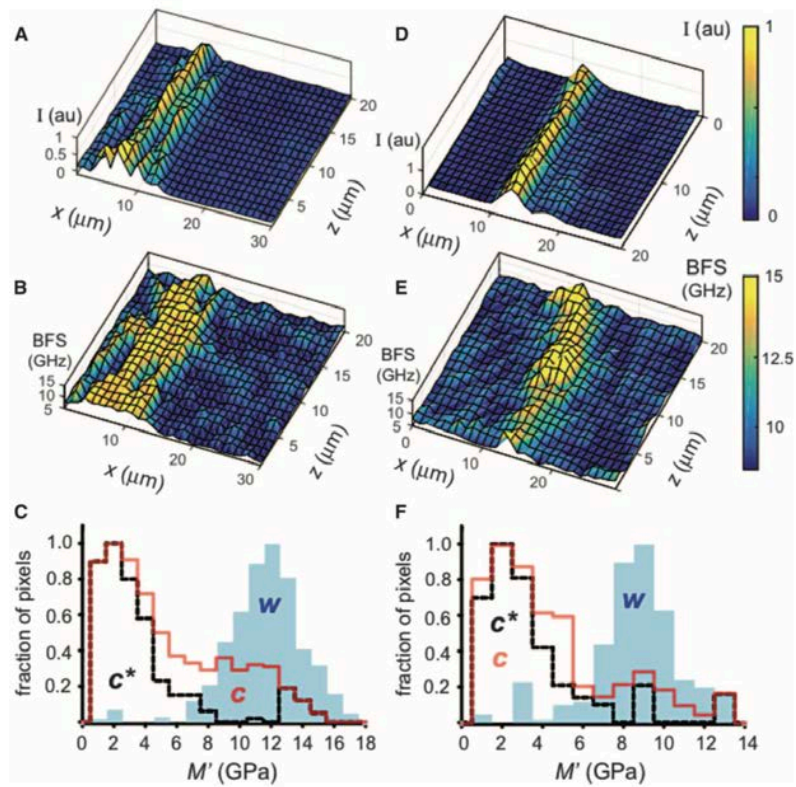

**Figure 17.** (A) Fluorescence intensity and (B) Brillouin maps of epidermal onion cells with membranes stained with a dye molecule. (C) Longitudinal elastic storage modulus M' (GPa) obtained by assigning extracellular matrix material properties to high fluorescence pixels (w, in light blue), cytoplasmic material properties to low fluorescence pixels (c, in red), and cytoplasmic material properties to all pixels at least 5 μm away from the plasma membrane (c*, in black). To avoid misinterpretation induced by possible dye diffusion, (D to F) report measurements similar to those in (A) to (C) but for dye molecule replaced by a plasma membrane marker expressed by a transgenic Arabidopsis hypocotyl cells. Figure reproduced from: ref. [92], American Association for the Advancement of Science.

Changes in cellular hydrostatic pressure were demonstrated to affect the stiffness of plant ECMs. FBi enabled the mechanical mapping of ECMs of the root, a plant organ that has so far not been amenable to mechanical characterization through perturbation-deformation methods due to its fragility.

The correlation between Brillouin frequency shift and plasma membrane fluorescence intensity was crucial to assess the overall contribution of the ECM to regions of increased stiffness, confirming the existence of a region of enhanced stiffness extending well beyond the apparent ECM in both onion and Arabidopsis epidermal cells deeper within the tissue.

## 5. FURTHER DEVELOPMENTS AND CHALLENGES

Despite the enabling capabilities of Brillouin spectroscopy and imaging, there are remaining challenges and opportunities for future directions in the fields of life sciences, medical imaging and healthcare technologies.

The most important aspect concerns the *qualitative* analysis of Brillouin scattering spectra. This truly mechanical fingerprint of materials contains the whole information 'compressed' within a single Brillouin peak. Whilst this vastly simplifies the analysis (only a curve fit to a single peak function is generally required), in heterogeneous samples the peak is an envelope of phonon resonances occurring at slightly different frequencies that are strongly overlapped. We first attempted to solve this issue using multivariate statistics,[27] which is generally applied to vibrational (IR and Raman) spectra but not to Brillouin spectra whereby modes are collective. The challenges posed by the spatial scale of these modes compared to the wavelength of light used to interrogate them and to the scale imposed by the optical setup, are to be taken into consideration (see Section 1.2) for an accurate description of Brillouin data.

The second equally important aspect concerns the *quantitative* analysis of Brillouin spectra. It has been shown that the intensity of Brillouin peaks correlates with the Raman intensity measured from the same location,[76] hence it is related to the system's polarizability anisotropy. Water is essentially isotropic, hence it has a weak scattering cross-section, making Raman and Brillouin spectroscopy suitable techniques for the study of live cells and tissues. On the other hand, the mechanical properties probed by Brillouin spectroscopy are strongly dependent on hydration and on relaxation dynamics in the GHz regime. Hence temporal scales (see Section 1.1), as well as spatial scales, need to be considered in the analysis of Brillouin data.

The third aspect is the relationship between the Brillouin frequency shift and the longitudinal elastic modulus at GHz frequency. This latter is different from the Young's modulus, despite some empirical correlations have been shown (e.g. log-log plot[35]). And it is not only a problem of spatial scale or frequency scale, which could be solved through multiscale theories of biomechanics and viscoelasticity. It is worth emphasising though the importance of developing theories that encompass multiple scales and of conducting computational studies at various scales, from molecules to larger assemblies. All these information are key to investigate the mechanical aspects of Brillouin data from complex materials. The comparison between different combinations of elastic moduli requires understanding of the underlying physics and biological system.

The considerations above lead to conclude that it is fundamentally important to be able to apply correlative techniques to Brillouin microscopy, and to provide additional information on the material. For instance, we have developed the combination of Brillouin and Raman microscopy to identify the molecular origin of the acoustic signals, and hence to investigate the contribution from various constituents e.g. water to the observed peaks.

However, equally important would be to be able to couple the Brillouin technique to methods that are capable to map the density and refractive index on a micrometric scale in order to extract stiffness maps from maps of $\omega_B$ (Eq.8a). Some progress has recently been reported on measuring dry masses via quantitative phase imaging.[93] A major concern about the use of this technique for our purposes is that it is based on phase imaging, the same method used to assess the refractive index. We feel that a real progress towards the evaluation of the ratio $\rho/n^2$ would be achieved by development of two independent methods for the local measure of $\rho$ and $n$.

Brillouin microscopy has been applied in combination with Optical Coherence Tomography (Br-OCT) to map the stiffness of a developing murine embryo while assessing structural changes.[94] Structural imaging by OCT enables to recognise developing organs, whilst Brillouin microscopy gains spatially resolved stiffness through the measure of the mean Brillouin shift. It can be anticipate this to be an area of further development in multimodal imaging.

Developmental biology and regenerative medicine are areas of major interest in current research in physics of life. Applications of Brillouin microscopy in the field of developmental biology include the work by Czarske and Guck on spinal cord development and repair in live zebrafish larvae,[95] whereby Brillouin scattering provides an optical micro-elastography method to assess tissue mechanics *in vivo*.

### 5.1. Multiscale Mechanics

Multiscale mechanics of biological systems is an emerging area of research and combines interdisciplinary approaches to basic research and technological advances. Brillouin microspectroscopy enters this realm as it provides a novel experimental tool for exploring the micromechanical properties of biological soft matter, including fibrous proteins and networks, cells and tissues. Brillouin spectroscopy techniques do not require applied load, hence they are free from aberrations and extrinsic probes. The inherent mechanical "proxy" are resonances from acoustic modes at high frequency. Through the Horizon2020 COST BioBrillouin network, we are currently exploring the possibilities of investigating biological systems, which span multiple scales, both spatial and temporal, and the challenges involved in bringing this knowledge together into a single multiscale understanding. We encourage the development of new theories where knowledge of the physics at each length scale can result in novel approaches to solving clinical challenges. This would stimulate further progress in biomedical photonics as it would bridge gaps that we currently recognise as limiting our capabilities in the field.

### 5.2. Coherent Nonlinear Methods

Stimulated Brillouin scattering (SBS) microscopy[96] imaging (BISTRO) has been shown by Meng et al. applied to simple liquids in quartz microchannels.[97]

Remer & Bilenca have developed a high-speed SBS spectrometer at 780 nm, with 10-100 fold increase in acquisition rate over frequency-domain SBS spectrometers.[42] This has been demonstrated in applications to water and Intralipid solutions.

## 6. CONCLUSIONS

Acoustic spectroscopy based on Brillouin scattering provides a detailed fingerprint of a material's micromechanics. Microscopy and imaging tools based on this technology have the potential to transform the fields of mechanobiology and clinical diagnostics. Recent technological advances have led to a renovated interest in Brillouin scattering from biological and clinical samples. Despite these strong drivers, there are fundamental scientific and technological challenges that need overcoming to enable further progress in biomedical science applications. In this review, we have examined the background theory and the state of the art of instrumentation and biomedical applications of Brillouin spectroscopy. We have also presented the outstanding challenges and progress towards translation, highlighting specific examples in the areas of single cell, cell population and tissue applications, from *in vitro* to *in vivo*. Both enabling capabilities and arising hurdles have been identified, that we hope will stimulate further advances in this fast moving field. Emerging consensus analysis was given, and the future perspectives of the field were assessed, in the context of national and international collaborative research initiatives, such as the Italian Society for Pure and Applied Biophysics and the EU COST Action BioBrillouin.

## ACKNOWLEDGMENTS

This work was supported by Cancer Research UK and the UK Engineering and Physical Sciences Research Council (NS/A000063/1; EP/M028739/1). The authors would like to thank the many co-workers and collaborators for their dedicated work and inspiring discussions to continuously advance this exciting research area.
## REFERENCES

(1) Brillouin, L. Diffusion de la lumière et des rayonnes X par un corps transparent homogène; influence de l'agitation thermique. *Ann. Phys.* **1922**, *17*, 88.
(2) Mandelstam, L. I. *J. Russ. Phys.-Chem. Soc.* **1926**, *58*, 831.
(3) Gross, E. Change of Wave-length of Light due to Elastic Heat Waves at Scattering in Liquids. *Nature* **1930**, *126*, 201.
(4) Bottani, C. E.; Fioretto, D. Brillouin scattering of phonons in complex materials. *Advances in Physics: X* **2018**, *3* (1), 1467281.
(5) Montrose, C. J.; Solovyev, V. A.; Litovitz, T. A. Brillouin Scattering and Relaxation in Liquids. *The Journal of the Acoustical Society of America* **1968**, *43* (1), 117.
(6) Comez, L.; Masciovecchio, C.; Monaco, G.; Fioretto, D. In *Solid State Physics*; Robert, E. C.;Robert, L. S., Eds.; Academic Press, 2012; Vol. 63.
(7) Mattana, S.; Mattarelli, M.; Urbanelli, L.; Sagini, K.; Emiliani, C.; Dalla Serra, M.; Fioretto, D.; Caponi, S. Non-contact mechanical and chemical analysis of single living cells by microspectroscopic techniques. *Light: Science & Applications* **2018**, *7*, e17139.
(8) Palombo, F.; Winlove, C. P.; Edginton, R. S.; Green, E.; Stone, N.; Caponi, S.; Madami, M.; Fioretto, D. Biomechanics of fibrous proteins of the extracellular matrix studied by Brillouin scattering. *J. R. Soc. Interface* **2014**, *11* (101), 12.
(9) Mattana, S.; Caponi, S.; Tamagnini, F.; Fioretto, D.; Palombo, F. Viscoelasticity of amyloid plaques in transgenic mouse brain studied by Brillouin microspectroscopy and correlative Raman analysis. *Journal of Innovative Optical Health Sciences* **2017**, *10* (5), 14.
(10) Palombo, F.; Madami, M.; Stone, N.; Fioretto, D. Mechanical mapping with chemical specificity by confocal Brillouin and Raman microscopy. *Analyst* **2014**, *139* (4), 729.
(11) Comez, L.; Fioretto, D.; Scarponi, F.; Monaco, G. Density fluctuations in the intermediate glassformer glycerol: A Brillouin light scattering study. *The Journal of Chemical Physics* **2003**, *119* (12), 6032.
(12) Janmey, P. A.; Hvidt, S.; Lamb, J.; Stossel, T. P. Resemblance of actin-binding protein/actin gels to covalently crosslinked networks. *Nature* **1990**, *345*, 89.
(13) Tempel, M.; Isenberg, G.; Sackmann, E. Temperature-induced sol-gel transition and microgel formation in \ensuremath{\alpha}-actinin cross-linked actin networks: A rheological study. *Phys. Rev. E* **1996**, *54* (2), 1802.
(14) Scopigno, T.; Yannopoulos, S. N.; Scarponi, F.; Andrikopoulos, K. S.; Fioretto, D.; Ruocco, G. Origin of the $\ensuremath{\lambda}$ Transition in Liquid Sulfur. *Phys. Rev. Lett.* **2007**, *99* (2), 025701.
(15) Rigato, A.; Miyagi, A.; Scheuring, S.; Rico, F. High-frequency microrheology reveals cytoskeleton dynamics in living cells. *Nature Physics* **2017**, *13*, 771.
(16) Fabry, B.; Maksym, G. N.; Butler, J. P.; Glogauer, M.; Navajas, D.; Fredberg, J. J. Scaling the Microrheology of Living Cells. *Phys. Rev. Lett.* **2001**, *87* (14), 148102.
(17) Sollich, P.; Lequeux, F.; Hébraud, P.; Cates, M. E. Rheology of Soft Glassy Materials. *Phys. Rev. Lett.* **1997**, *78* (10), 2020.
(18) MacKintosh, F. C.; Käs, J.; Janmey, P. A. Elasticity of Semiflexible Biopolymer Networks. *Phys. Rev. Lett.* **1995**, *75* (24), 4425.
23